\documentclass[%
superscriptaddress,
showpacs,
nofootinbib,
amsmath,amssymb,
prc,
twocolumn,
floatfix ]%
{revtex4-1}

\usepackage{color}
\usepackage{CJK}

\usepackage{graphicx}
\usepackage{dcolumn}
\usepackage{bm}
\usepackage{hyperref}

\allowdisplaybreaks

\begin{document}

\begin{CJK*}{UTF8}{}
\title{Time-dependent generator coordinate method study of fission: dissipation effects}
\CJKfamily{gbsn}
\author{Jie Zhao (赵杰)}%
\affiliation{Center for Circuits and Systems, Peng Cheng Laboratory, Shenzhen 518055, China}
\author{Tamara Nik\v{s}i\'c}%
\affiliation{Physics Department, Faculty of Science, University of Zagreb, Bijeni\v{c}ka Cesta 32,
        	      Zagreb 10000, Croatia}             
\author{Dario Vretenar}%
\affiliation{Physics Department, Faculty of Science, University of Zagreb, Bijeni\v{c}ka Cesta 32,
              Zagreb 10000, Croatia}
 \affiliation{ State Key Laboratory of Nuclear Physics and Technology, School of Physics, Peking University, Beijing 100871, China}            

\date{\today}

\begin{abstract}
Starting from a quantum theory of dissipation for nuclear collective motion, the time-dependent generator coordinate method (TDGCM) is extended to allow for dissipation effects in the description of induced fission dynamics. The extension is based on a generalization of the GCM generating functions that includes excited states, and the resulting equation of motion in the collective coordinates and excitation energy. With the  assumption of a narrow hamiltonian kernel, an expansion in a power series in collective momenta leads to a Schr\"odinger-like equation that explicitly includes a dissipation term, proportional to the momentum of the statistical wave function.  An illustrative calculation is performed for induced fission of $^{228}$Th. The three-dimensional model space includes the axially-symmetric quadrupole and octupole shape variables, and the nuclear temperature. When compared to data for photo-induced fission of $^{228}$Th, the calculated fission yields demonstrate the important role of the additional term in the hamiltonian that explicitly takes into account the dissipation of energy of collective motion into intrinsic degrees of freedom.
\end{abstract}

\maketitle

\end{CJK*}

\bigskip

\section{Introduction~\label{sec:Introduction}}
Among diverse methods that have been developed over many decades to describe the dynamics of low-energy induced fission, the time-dependent generator coordinate method (TDGCM) \cite{krappe12,schunck16,younes19,bender20,verriere20} presents the only microscopic approach that can be  used to model the entire fission process, from the quasi-stationary initial state all the way to scission and the emergence of fission fragments. A recent implementation of the TDGCM \cite{Regnier2016_CPC200-350,Regnier2018_CPC225-180}, based on the Gaussian overlap approximation, has been successfully employed in a number of fission studies \cite{Regnier16,Regnier19,Tao2017_PRC96-024319,Zhao2019_PRC99-014618,Zhao2019_PRC99-054613,Zhao2020_PRC101-064605,Zhao2021_PRC104-044612,Verriere21} 
that have explored many interesting subjects, such as the influence of static pairing correlations on fission yields, the definition of scission configurations, 
different approximations for the collective inertia tensor, 
finite temperature effects, the role of dynamical pairing degrees of freedom, and symmetry restoration. 

TDGCM presents a fully quantum mechanical approach that describes fission dynamics as an adiabatic evolution of collective degrees of freedom. The nuclear wave function is represented by a linear superposition of many-body generator states that are functions of collective coordinates (shape parameters, pairing degrees of freedom). By employing the Gaussian overlap approximation (GOA), the GCM Hill-Wheeler equation of motion for the collective wave function reduces to a local, time-dependent Schr\"odinger equation in the space of collective coordinates. Given an energy density functional (EDF) or effective interaction, the collective potential and inertia tensor are determined by constrained self-consistent mean-field calculations in the space of collective coordinates. The TDGCM+GOA is particularly suitable for the slow evolution from the quasi-stationary initial state to the outer fission barrier (saddle point). Beyond the saddle point fission dynamics becomes dissipative as the nucleus elongates towards scission. The effect of dissipation is to heat the system, modify the path in the multidimensional deformation space, increase the time to scission, and generate fluctuations in various quantities \cite{nix83}. However, 
in contrast to models based on time-dependent density functional theory (TDDFT)  \cite{simenel12,simenel18,nakatsukasa16,stevenson19,bulgac16,magierski17,scamps18,bulgac19,bulgac20}, TDGCM does not contain a one-body dissipation mechanism, because it only takes into account collective degrees of freedom in the adiabatic approximation.

Since dissipation of the energy of collective motion into intrinsic degrees of freedom plays an important role in the dynamics of the final stage of the fission process, it is important to extend the adiabatic TDGCM approach to explicitly include nucleon degrees of freedom. This is, of course, very difficult to achieve in a fully microscopic approach. In the Schr\"odinger collective-intrinsic model (SCIM) \cite{bernard11,younes19}, the coupling between collective and intrinsic excitations is taken into account by a generalization of the GCM based on zero- and two-quasiparticle excitations. Another interesting microscopic approach is the transport theory of Ref.~\cite{dietrich10}, in which the slow evolution of the nuclear shape is treated explicitly, while the fast time-dependent intrinsic excitations (multi-quasiparticle states) are described in a statistical approximation. This model is also based on deformation constrained self-consistent mean-field calculations and a generalization of the GCM. The TDGCM extensions of Refs.~\cite{bernard11,younes19} and \cite{dietrich10} both have solid microscopic foundations, but they are also extremely complex, and have yet to be implemented in a realistic calculation of fission dynamics. 

In the present study we revive a microscopic theory of dissipation for nuclear collective motion, introduced by Kerman and Koonin in Ref.~\cite{Kerman1974_PS10-118}. Based on a generalization of the GCM generating functions to include excited intrinsic states, and certain statistical assumptions, a quantal equation of motion was derived in both the collective coordinates and excitation energy. With the usual assumption of a narrow hamiltonian kernel, a Schr\"odinger-like equation can be derived for the statistical collective wave function, including dissipation. In Ref.~\cite{Kerman1974_PS10-118} the method was illustrated with a simple one-dimensional model calculation of heavy-ion collision. Here we employ this method to extend our implementation of the temperature-dependent TDGCM for induced fission dynamics, to include dissipation effects in the collective space of axial quadrupole and octupole deformations. The formalism is developed in Sec.~\ref{sec:model}. In Sec.~\ref{sec:results} we present an illustrative calculation of charge yields for induced fission of $^{228}$Th. Finally, Section~\ref{sec:summary} contains a short summary and an outlook for future studies. 

\section{\label{sec:model}Theoretical framework}
The purpose of the present study is to extend the time-dependent generator coordinate method (TDGCM) by including energy dissipation, and apply the model to the  
description of induced fission dynamics. The method is based on the quantum theory of dissipation for nuclear collective motion of Ref.~\cite{Kerman1974_PS10-118}.

The derivation starts with a trial TDGCM many-body wave function of the following form: 
\begin{equation}
\label{eq:wavefunction}
\left| \Phi(t) \right\rangle = \sum_{n} \int d\bm{q} f_{n} (\bm{q}, t) \left| n\bm{q} \right\rangle,
\end{equation}
where $\bm{q}$ denotes the set of collective coordinates, $n$ labels the excited states at each value of the collective coordinate $\bm{q}$, and $f_{n} (\bm{q}, t)$  are weight functions.
From the time-dependent variational principle 
\begin{equation}
\label{eq:variation}
\delta \int \left\langle \Phi(t) \left| \left[ \hat{H} - i\hbar \partial_t \right] \right| \Phi(t) \right\rangle dt = 0,
\end{equation}
the matrix integral Hill-Wheeler equation is obtained
\begin{equation}
\label{eq:tdgcm}
\begin{aligned}
& \sum_{n^{\prime}} \int d\bm{q}^{\prime} \left\{ 
	\mathcal{H}_{nn^{\prime}} (\bm{q}, \bm{q}^{\prime}) f_{n^{\prime}} (\bm{q}^{\prime}, t) \right. \\
	 & \left. - \mathcal{N}_{nn^{\prime}} (\bm{q}, \bm{q}^{\prime}) 
	\left[ i\hbar\partial_t f_{n^{\prime}} (\bm{q}^{\prime}, t) \right]
\right\} = 0,
\end{aligned}
\end{equation}
where $\mathcal{H}_{nn^{\prime}} (\bm{q}, \bm{q}^{\prime}) = \left\langle n\bm{q} \left| \hat{H} \right| n^{\prime}\bm{q}^{\prime} \right\rangle$ is the Hamiltonian kernel, 
and $ \mathcal{N}_{nn^{\prime}} (\bm{q}, \bm{q}^{\prime}) = \left\langle \left. n\bm{q} \right| n^{\prime}\bm{q}^{\prime} \right\rangle$ is
the norm overlap kernel.
It is useful to express Eq.~(\ref{eq:tdgcm}) in terms of another set of functions $g_{n} (\bm{q}, t)$, defined by
\begin{equation}
\label{eq:reduced_f}
g_{n} (\bm{q}, t) = \sum_{n^{\prime}} \int d\bm{q}^{\prime}
	\mathcal{N}_{nn^{\prime}}^{1/2} (\bm{q}, \bm{q}^{\prime}) f_{n^{\prime}} (\bm{q}^{\prime}, t).
\end{equation}
With this transformation, Eq.~(\ref{eq:tdgcm}) takes the form
\begin{equation}
\label{eq:reduced_tdgcm}
i\hbar \partial_t g_{n} (\bm{q}, t) = \sum_{n^{\prime}} \int d\bm{q}^{\prime} 
	H_{nn^{\prime}} (\bm{q}, \bm{q}^{\prime}) g_{n^{\prime}} (\bm{q}^{\prime}, t),
\end{equation}
with 
\begin{equation}
\label{eq:reduced_h}
\begin{aligned}
H_{nn^{\prime}} (\bm{q}, \bm{q}^{\prime}) =& \sum_{n_{1} n_{2}} \int d\bm{q}_{1} d\bm{q}_{2} 
	  \mathcal{N}_{nn_{1}}^{-1/2} (\bm{q}, \bm{q}_{1}) \\
	  &\times \mathcal{H}_{n_{1}n_{2}} (\bm{q}_{1}, \bm{q}_{2}) 
	  \mathcal{N}_{n_{2}n^{\prime}}^{-1/2} (\bm{q}_{2}, \bm{q}^{\prime}).
\end{aligned}
\end{equation}

As noted in Ref.~\cite{Kerman1974_PS10-118}, 
the level density for each value of the collective coordinate $\bm{q}$ is high even at relatively low excitation energies, so that  
the discrete label $n$ can be separated into a continuous excitation energy variable $\epsilon$, and a degeneracy label $\lambda$:
\begin{equation}
\label{eq:ld}
\sum_{\lambda, ~\rm{fixed} ~\epsilon} = \rho(\bm{q}, \epsilon) d\epsilon,
\end{equation}
where $\rho(\bm{q}, \epsilon)$ denotes the level density at the point 
$\bm{q}$ in the collective coordinate space. 
We can then substitute  
\begin{equation}
\label{eq:wfl}
g_{n} (\bm{q},t) \rightarrow g_{\lambda} (\bm{q}, \epsilon; t), 
\end{equation}
\begin{equation}
\label{eq:hll}
H_{nn^{\prime}} (\bm{q}, \bm{q}^{\prime}) \rightarrow H_{\lambda \lambda^{\prime}} (\bm{q}, \bm{q}^{\prime}; \epsilon, \epsilon^{\prime}),
\end{equation}
and rewrite Eq. (\ref{eq:reduced_tdgcm})  as 
\begin{equation}
\label{eq:tdgcm_e}
\begin{aligned}
& i\hbar \partial_t g_{\lambda} (\bm{q}, \epsilon; t) = 
	\int d\bm{q}^{\prime} H_{\lambda \lambda} (\bm{q}, \bm{q}^{\prime}; \epsilon, \epsilon) g_{\lambda} (\bm{q}^{\prime}, \epsilon; t) \nonumber \\
	&+ \sum_{\lambda^{\prime} \neq \lambda} \int \int d\bm{q}^{\prime} d\epsilon^{\prime}
	H_{\lambda \lambda^{\prime}} (\bm{q}, \bm{q}^{\prime}; \epsilon, \epsilon^{\prime}) g_{\lambda^{\prime}} (\bm{q}^{\prime}, \epsilon^{\prime}; t).
\end{aligned}
\end{equation}

Following the prescription of Ref.~\cite{Kerman1974_PS10-118}, $g_{\lambda} (\bm{q}, \epsilon; t)$ in Eq.~(\ref{eq:tdgcm_e}) is replaced by its average value $\overline{g_{\lambda} (\bm{q}, \epsilon; t)}$, 
and the statistical wave function $\psi (\bm{q}, \epsilon; t)$ is defined as 
\begin{equation}
\label{eq:swf}
\overline{g_{\lambda} (\bm{q}, \epsilon; t)} = { \psi (\bm{q}, \epsilon; t) \over \sqrt{\rho(\bm{q}, \epsilon)} }.
\end{equation}
Performing the summation of Eq.~(\ref{eq:tdgcm_e}) over $\lambda$, one obtains
\begin{equation}
\label{eq:tdgcm_f}
\begin{aligned}
& i\hbar \frac{\partial}{\partial t} \psi (\bm{q}, \epsilon; t) =
	\int d\bm{q}^{\prime} h(\bm{q}, \bm{q}^{\prime}; \epsilon, \epsilon) \psi(\bm{q}^{\prime}, \epsilon; t) \\
	& + \sum_{\lambda^{\prime} \neq \lambda} \int \int d\bm{q}^{\prime} d\epsilon^{\prime} 
	 h(\bm{q}, \bm{q}^{\prime}; \epsilon, \epsilon^{\prime}) \psi(\bm{q}^{\prime}, \epsilon^{\prime}; t),
\end{aligned}
\end{equation}
with the statistical Hamiltonian kernels are defined as 
\begin{equation}
\label{eq:hdiag}
h(\bm{q}, \bm{q}^{\prime}; \epsilon, \epsilon) = {1 \over \sqrt{\rho(\bm{q}, \epsilon)} } 
	\sum_{\lambda} H_{\lambda \lambda} (\bm{q}, \bm{q}^{\prime}; \epsilon, \epsilon) 
	{1 \over \sqrt{\rho(\bm{q}^{\prime}, \epsilon)} },
\end{equation}
\begin{equation}
\label{eq:h_nondiag}
h(\bm{q}, \bm{q}^{\prime}; \epsilon, \epsilon^{\prime}) = {1 \over \sqrt{\rho(\bm{q}, \epsilon)} } 
	\sum_{\lambda, \lambda^{\prime} \neq \lambda} H_{\lambda \lambda^{\prime}} (\bm{q}, \bm{q}^{\prime}; \epsilon, \epsilon^{\prime}) 
	 {1 \over \sqrt{\rho(\bm{q}^{\prime}, \epsilon^{\prime})} }.
\end{equation}

The usual GCM assumption that the Hamiltonian overlap kernel decreases rapidly with increasing 
$\left| \bm{q} - \bm{q}^{\prime} \right|$ (in comparison to the scale of 
variations in the statistical wave function $\psi$), enables an expansion of Eq.~(\ref{eq:tdgcm_f}) in a power series in collective momenta $\bm{P}={-}i\hbar (\partial / \partial\bm{q})$, leading to a Schr{\"o}dinger-like equation
\begin{equation}
\label{eq:tdgcmgoa}
\begin{aligned}
i\hbar \partial_t \psi (\bm{q}, \epsilon; t)
	=& \left[ h^{(0)}(\bm{q}; \epsilon)
		- \frac{1}{8\hbar^{2}} \left[\bm{P}^{2} h^{(2)}(\bm{q}; \epsilon) \right] \right. \\
		& \left. - \frac{1}{2\hbar^{2}} \bm{P} h^{(2)}(\bm{q}; \epsilon) \bm{P} \right] \psi (\bm{q}, \epsilon; t) \\
		&+ \frac{i}{2\hbar} \int \left\{ \bm{P}, h^{(1)}(\bm{q}; \epsilon, \epsilon^{\prime}) \right\} \psi (\bm{q}, \epsilon^{\prime}; t) d\epsilon^{\prime},
\end{aligned}
\end{equation}
where 
\begin{equation}
\label{eq:h_n}
h^{(n)} (\bm{q}; \epsilon, \epsilon^{\prime}) = \int \bm{z}^{n} h(\bm{q}+\bm{z}/2, \bm{q}-\bm{z}/2; \epsilon, \epsilon^{\prime}) d\bm{z},
\end{equation}
$\bm{z} = \bm{q} - \bm{q}^{\prime}$, and $h^{(n)} (\bm{q}; \epsilon) \equiv h^{(n)} (\bm{q}; \epsilon, \epsilon^{\prime})$.
The curly brackets in the last term of Eq.~(\ref{eq:tdgcmgoa}) denote anti-commutation. With the collective potential defined as 
\begin{equation}
\label{eq:potential}
V(\bm{q}, \epsilon) = h^{(0)}(\bm{q}; \epsilon) - \frac{1}{8\hbar^{2}} \left[\bm{P}^{2} h^{(2)}(\bm{q}; \epsilon) \right],
\end{equation}
and the collective mass 
\begin{equation}
\label{eq:mass}
\mathcal{M}(\bm{q}, \epsilon) = - \frac{1}{\hbar^{2}} h^{(2)}(\bm{q}; \epsilon),
\end{equation}
Eq.~(\ref{eq:tdgcmgoa}) can be written in the final form
\begin{equation}
\label{eq:tdgcmgoa_f}
\begin{aligned}
i\hbar \partial_t \psi (\bm{q}, \epsilon; t) 
	=& \left[ V(\bm{q}, \epsilon)
		+ \bm{P} \frac{1}{2 \mathcal{M}(\bm{q}, \epsilon)} \bm{P} \right] \psi (\bm{q}, \epsilon; t) \\
		&+ \frac{i}{2} \int \left\{ \bm{P}, \bm{\eta}(\bm{q}; \epsilon, \epsilon^{\prime}) \right\} \psi (\bm{q}, \epsilon^{\prime}; t) d\epsilon^{\prime}.
\end{aligned}
\end{equation}
The dissipation function $\bm{\eta}(\bm{q}; \epsilon, \epsilon^{\prime}) = h^{(1)}(\bm{q}; \epsilon, \epsilon^{\prime}) / \hbar$ is 
anti-hermitian in the variables $\epsilon$ and $\epsilon^{\prime}$, so that the Hamiltonian is still a hermitian operator.
For further details we refer the reader to Ref.~\cite{Kerman1974_PS10-118}.

In practical applications, the excitation energy $\epsilon$ is a function of temperature, and  
the energy dependent collective potential $V(\bm{q}, \epsilon)$, mass $\mathcal{M}(\bm{q}, \epsilon)$, and dissipation function $\bm{\eta}(\bm{q}; \epsilon, \epsilon^{\prime})$ can be expressed in terms of the nuclear temperature $T$:
$V(\bm{q}, T) \equiv V(\bm{q}, \epsilon(T))$, $\mathcal{M}(\bm{q}, T) \equiv \mathcal{M}(\bm{q}, \epsilon(T))$,
and $\eta(\bm{q}; T, T^{\prime}) \equiv \eta(\bm{q}; \epsilon(T), \epsilon(T^{\prime}))$.
Therefore, Eq.~(\ref{eq:tdgcmgoa_f}) takes the form:
\begin{equation}
\label{eq:tdgcmgoa_ft}
\begin{aligned}
i\hbar \partial_t \psi (\bm{q}, T; t)
	=& \left[ V(\bm{q}, T)
		+ \bm{P} \frac{1}{2 \mathcal{M}(\bm{q}, T)} \bm{P} \right] \psi (\bm{q}, T; t) \\
		&+ \frac{i}{2} \int \left\{ \bm{P}, \bm{\mathcal{O}}(\bm{q}; T, T^{\prime}) \right\} \psi (\bm{q}, T^{\prime}; t) dT^{\prime},
\end{aligned}
\end{equation}
with $\bm{\mathcal{O}}(\bm{q}; T, T^{\prime}) = \bm{\mathcal{\eta}}(\bm{q}; T, T^{\prime}) d\epsilon(T) / dT$.

In the present study we employ the self-consistent multidimensionally-constrained (MDC) relativistic Hartree-Bogoliubov (RHB) 
model ~\cite{Lu2014_PRC89-014323,Zhao2017_PRC95-014320} at finite
temperature~\cite{Goodman1981_NPA352-30,Egido1986_NPA451-77,Zhao2020_PRC101-064605}. 
In a grand-canonical ensemble, the expectation value of an operator $\hat{O}$ is given by the ensemble average
\begin{equation}
\label{eq:O_expectation}
\langle \hat{O} \rangle = \textnormal{Tr}\left[ \hat{D}\hat{O}  \right],
\end{equation}
where $\hat{D}$ is the density operator
\begin{equation}
\label{eq:density_operator}
\hat{D} = \frac{1}{Z} e^{-\beta\left( \hat{H}-\mu \hat{N}\right)}.
\end{equation}
$Z$ is the partition function, $\beta=1/k_BT$ is the inverse temperature with the Boltzmann constant $k_B$. $\hat{H}$ is the Hamiltonian
of the system, $\mu$ denotes the chemical potential, and $\hat{N}$ is the particle number operator. 
The entropy of the nuclear system is $S=-k_B\langle \hat{D}\ln{\hat{D}}\rangle$. The intrinsic level density $\rho$ is calculated
in the saddle-point approximation \cite{Bohr1969_Nucl_Structure}
\begin{equation}
\label{eq:level_density}
\rho = \frac{e^S}{(2\pi)^{3/2}D^{1/2}},
\end{equation}
where $D$ is the determinant of a $3\times 3$ matrix that contains the second derivatives of the entropy with respect to $\beta$ and
$\nu_\tau=\beta \mu_\tau$ $(\tau \equiv p, n )$ at the saddle point.

 The finite temperature relativistic Hartree-Bogoliubov 
(FT-RHB) equations are obtained by minimizing the grand-canonical potential $\Omega =\langle \hat{H}\rangle +TS -\mu_\tau\langle\hat{N}\rangle$. In this work 
the particle-hole channel is specified by the choice of the relativistic energy density functional DD-PC1~\cite{Niksic2008_PRC78-034318}, 
while pairing correlations are taken into account in the Bardeen-Cooper-Schrieffer (BCS) approximation with a finite-range separable 
pairing force~\cite{Tian2009_PLB676-44}. 
The parameters of the pairing interaction have been adjusted to reproduce the empirical pairing gaps 
in the mass region considered in this study~\cite{Zhao2019_PRC99-054613}. 
The nuclear shape is parameterized by the deformation parameters
\begin{equation}
 \beta_{\lambda\mu} = {4\pi \over 3AR^\lambda} \langle Q_{\lambda\mu} \rangle.
\end{equation}
The shape is assumed to be invariant under the exchange of the $x$ and $y$ axes, 
and all deformation parameters $\beta_{\lambda\mu}$ with even $\mu$ can be included simultaneously.
The self-consistent relativistic mean-field (RMF+BCS) equations are solved by an expansion in the 
axially deformed harmonic oscillator (ADHO) basis~\cite{Gambhir1990_APNY198-132}.
In the present study calculations have been performed 
in an ADHO basis truncated to $N_f = 20$ oscillator shells.

The internal excitation energy $\epsilon(T)$ of a nucleus at temperature $T$ is defined as the difference between the total binding energy 
of the equilibrium RMF+BCS minimum at temperature $T$ and at $T=0$.
The thermodynamical potential relevant for deformation effects is the Helmholtz free energy $F(T) = E(T) - TS$, evaluated at constant temperature $T$, 
where $E(T)$ is the binding energy of the deformed nucleus, and the deformation-dependent energy landscape is obtained in a self-consistent
finite-temperature mean-field calculation with constraints on the mass multipole moments $Q_{\lambda\mu} = r^{\lambda} Y_{\lambda\mu}$.

In the present case the collective coordinates $\bm{q}$ correspond to the quadrupole $\langle Q_{20} \rangle$ and 
octupole  $\langle Q_{30} \rangle$ mass multipole moments.
The collective potential is, therefore, $V(\bm{q}, T) = \epsilon(T) + F(\bm{q}, T)$, where $F(\bm{q}, T)$ is the Helmholtz free energy
normalized to the corresponding value at the equilibrium RMF+BCS minimum at temperature $T$. 
The mass tensor $\mathcal{M}(\bm{q}, T)$ is calculated 
in the finite-temperature perturbative cranking approximation~\cite{Zhu2016_PRC94-024329,Martin2009_IJMPE18-861}
\begin{equation}
\label{eq:pmass}
\mathcal{M}^{Cp} = \hbar^2 {\it M}_{(1)}^{-1} {\it M}_{(3)} {\it M}_{(1)}^{-1}, 
\end{equation}
with
\begin{widetext}
\begin{equation}
\begin{aligned}
 \left[ M_{(k)}\right]_{ij,T} =& \frac{1}{2} \sum_{\mu \neq \nu} 
	      \langle 0 | \hat{Q}_{i} | \mu \nu \rangle
	     \langle \mu \nu | \hat{Q}_{j} | 0 \rangle 
	     \left\{ { (u_{\mu} u_{\nu} - v_{\mu} v_{\nu})^{2} \over (E_{\mu} - E_{\nu})^{k} } 
	      \left[ \tanh\left( {E_{\mu} \over 2k_{B}T}  \right) - \tanh\left( {E_{\nu} \over 2k_{B}T} \right) \right]  \right\}  \\
	&+ {1 \over 2} \sum_{\mu \nu} 
	     \langle 0 | \hat{Q}_{i} | \mu \nu \rangle
	     \langle \mu \nu | \hat{Q}_{j} | 0 \rangle 
	     \left\{ { (u_{\mu} v_{\nu} + u_{\nu} v_{\mu})^{2} \over (E_{\mu} + E_{\nu})^{k} } 
	     \left[ \tanh\left( {E_{\mu} \over 2k_{B}T}  \right) + \tanh\left( {E_{\nu} \over 2k_{B}T} \right) \right]  \right\}.
\label{eq:per-mass}
\end{aligned}
\end{equation}
\end{widetext}
$| \mu\nu \rangle$ are the two-quasiparticle states with the corresponding quasiparticle energies $E_{\mu}$ and $E_{\nu}$. $v_{\mu}^{2}$ are the BCS occupation probabilities, and $u_{\mu}^{2} = 1 - v_{\mu}^{2}$.

Equation (\ref{eq:tdgcmgoa_ft}) describes nuclear collective motion with dissipation. In addition to the non-dissipative potential and kinetic energy terms, the dissipative channel coupling is proportional to the momentum of the collective wave function. Even though the dissipation function $\bm{\eta}(\bm{q}; \epsilon, \epsilon^{\prime})$ could, in principle, be determined in a fully microscopic way, in practice this is extremely difficult. This is why, in an exploratory study, we will approximate the dissipation function with a phenomenological ansatz. 
As explained in Ref.~\cite{Kerman1974_PS10-118}, for complicated nuclear many-body configurations, 
the sign of the Hamiltonian kernel $h (\bm{q}, \bm{q}^{\prime}; \epsilon, \epsilon^{\prime})$  changes randomly with variation of the internal excitation energies
$\epsilon$ and $\epsilon^{\prime}$.
Following the central limit theorem, one would expect that the values of the dissipation function
$\bm{\mathcal{\eta}}(\bm{q}; T, T^{\prime})$ are random variables whose
probability density corresponds to a Gaussian distribution centered around zero.
Thus we choose the dissipation function $\bm{\mathcal{\eta}}(\bm{q}; T, T^{\prime})$ to be of the form
\begin{equation}
\label{eq:eta}
 \bm{\mathcal{\eta}}(\bm{q}; T, T^{\prime}) = 
 \begin{cases}
	 0 			             & \beta_{2}< \beta_{2}^{0}         \\
	 \bm{\eta}(T, T^\prime)  & \beta_{2} \geq \beta_{2}^{0},
 \end{cases} 
\end{equation}
where the matrix elements $\bm{\eta}(T, T^\prime)$ are Gaussian random variables. $\beta_{2}^{0}$ is set to $1.5$, which is slightly beyond the second fission barrier for the example of fission of $^{228}$Th, that will be considered in the next section. The cut-off value $\beta_2^0$
is introduced, because for induced nuclear fission one only expects significant dissipation effects in the saddle to scission phase. Similar to Ref.~\cite{Kerman1974_PS10-118}, 
the root-mean-square value of the Gaussian distribution of the $\bm{\eta}(T, T^\prime)$ random variables reads  $\displaystyle \gamma \sqrt{ \log[\rho(T)] \log[\rho(T^{\prime})] }$. 
In this expression $\rho(T)$ is the intrinsic nuclear level density calculated at the RMF+BCS equilibrium minimum, while $\gamma$ is an adjustable parameter. Note that in 
Ref.~\cite{Kerman1974_PS10-118} the ansatz $\displaystyle \gamma \sqrt{\rho(T) \rho(T^{\prime}) }$ was used. However, in the realistic example considered here the level density is much higher, and using the expression without the $\log$ functions leads to numerical instabilities. The precise value of $\gamma$  is not crucial but, of course, its strength must be such to produce a noticeable effect.

To model the dynamics of the fission process we follow 
 the time-evolution of an initial wave packet $g(\bm{q},T,t=0)$, built 
as a Gaussian superposition of quasi-bound states $g_k$
\begin{equation}
\label{eq:initial-wave-packet}
\psi(\bm{q},T,t=0) = \sum_k{e^{(E_k-\bar{E})/\left(2\sigma^2\right)}g_k(\bm{q},T)},
\end{equation}
where the value of the parameter $\sigma$ is set to 0.5 MeV. The collective states $\{g_k(\bm{q},T)\}$
are solutions of the stationary eigenvalue equation, in which the original collective potential
is replaced by a new potential $V^\prime(\bm{q},T)$ that is obtained by extrapolating the inner potential
barrier with a quadratic form. A more detailed description of this procedure can be found in Ref.~\cite{Regnier2018_CPC225-180}. 
The average energy of the collective initial state is calculated as
\begin{equation}
\label{eq:ecoll}
E_{\rm coll}^* = \left\langle \psi(\bm{q},T,t=0) \right| H_{\rm coll} \left| \psi(\bm{q},T,t=0)\right\rangle,
\end{equation}
where $H_{\rm coll}$ is the collective Hamiltonian (without dissipation), and the mean energy $\bar{E}$ in Eq.~(\ref{eq:initial-wave-packet}) is adjusted iteratively to obtain the chosen value of $E_{\rm coll}^*$.

The time-evolution is described by Eq.~(\ref{eq:tdgcmgoa_ft}), in which 
the temperature $T$ is treated as the third coordinate.
Following the prescription of Ref.~\cite{Regnier2016_CPC200-350,Regnier2018_CPC225-180}, 
we have discretized the three-dimensional (3D) space $(\bm{q},T)$ with the 
continuous Galerkin finite element method. This leads to 
a large set of coupled, time-dependent Schr{\"o}dinger-like equations characterized by sparse overlap 
and Hamiltonian matrices. The solution is evolved in small time steps by applying an explicit and unitary propagator built as a Krylov approximation of the exponential of the Hamiltonian. 
The time step is $\delta t=5\times 10^{-4}$ zs (1 zs $= 10^{-21}$ s), and the charge and mass 
distributions are calculated after $4\times10^{4}$ time steps, which correspond to 20 zs.
As in our recent calculations of Refs.~\cite{Tao2017_PRC96-024319,Zhao2019_PRC99-014618,Zhao2019_PRC99-054613,
Zhao2020_PRC101-064605,Zhao2021_PRC104-044612}, 
the parameters of the additional imaginary absorption potential that takes into account the escape 
of the collective wave packet  in the domain outside the region of calculation \cite{Regnier2018_CPC225-180} are: 
the absorption rate $r=20\times 10^{22}$ s$^{-1}$ and the width of the absorption band $w=1.0$.

The deformation collective space is divided into an inner region with a single nuclear density distribution, 
and an external region that contains the two separate fission fragments. 
The scission hyper-surface that divides the inner and external regions is determined by calculating the 
expectation value of the 
Gaussian neck operator $\displaystyle \hat{Q}_{N}=\exp[-(z-z_{N})^{2} / a_{N}^{2}]$, 
where $a_{N}=1$ fm and $z_{N}$ is the position of the neck~\cite{Younes2009_PRC80-054313}.
We define the pre-scission domain by $\langle \hat{Q}_{N} \rangle>3$, and consider the frontier of this domain as the scission surface.
 The flux of the probability current through this hyper-surface provides a measure of the probability of observing a 
 given pair of fragments at time $t$~\cite{Regnier2018_CPC225-180}.
 
 From Eq.~\ref{eq:tdgcmgoa_ft}, the time evolution of the probability density reads
 \begin{widetext}
\begin{equation}
\label{eq:probability}
\begin{aligned}
\frac{\partial}{\partial t} \left| \psi(\bm{q}, T; t) \right|^{2} = &-\sum_{i} { \partial J_{i}(\bm{q},T; t) \over \partial q_i} 
	  +\frac{i}{2} \left\{ 
		\psi^{*}(\bm{q}, T; t) \int dT^{\prime} \sum_{i} \left\{ \partial_{q_{i}}, \eta_{i} (\bm{q};T,T^{\prime}) \right\} \psi(\bm{q}, T^{\prime}; t)  \right. \\
		& \left. - \psi(\bm{q}, T; t) \int dT^{\prime} \sum_{i} \left\{ \partial_{q_{i}}, \eta_{i} (\bm{q};T,T^{\prime}) \right\} \psi^{*}(\bm{q}, T^{\prime}; t)
		\right\}.
\end{aligned}
\end{equation}
\end{widetext}
The two-dimensional scission hyper-surface is embedded in the three-dimensional space  $(\bm{q}, T)$. 
However, for each value of $T$ we have some value of shape coordinates
$\bf{q}$ which correspond to scission contour (here, $\langle \hat{Q}_{N} \rangle=3$) and, thus, by 
taking these values of the shape coordinates and integrating
along $T^\prime$ in the last term of Eq.~(\ref{eq:probability}), we would inevitably leave the scission contour. Hence, in calculations of the integrated flux through the scission hyper-surface we
neglect the last term of Eq.~(\ref{eq:probability}), and consider the current 
\begin{equation}
\label{eq:current}
J_{i}(\bm{q},T; t) = \hbar \sum_{j} \mathcal{M}_{ij}^{-1}(\bm{q},T) \rm{Im}\left( \psi^{*} \frac{\partial\psi}{\partial q_{j}} \right),
\end{equation}
 to obtain the usual continuity equation
\begin{equation}
\label{eq:continuity}
\frac{\partial}{\partial t} \left| \psi(\bm{q}, T; t) \right|^{2} = -\sum_{i} { \partial J_{i}(\bm{q},T; t) \over \partial q_i}.
\end{equation}
 The flux of the probability current $\bm{J}(\bm{q},T; t)$ through the scission hyper-surface provides a measure of the probability of observing a 
 given pair of fragments at time $t$. Each infinitesimal surface element is associated with a given pair of fragments 
 $(A_L,A_H)$, where $A_L$ and $A_H$ denote the lighter and heavier fragments, respectively. The integrated flux 
 $F(\xi;t)$ for a given surface element $\xi$ is defined as~\cite{Regnier2018_CPC225-180}
\begin{equation}
F(\xi;t) = \int_{t_0}^{t} dt^\prime \int_{(\bm{q},T) \in \xi} \bm{J}(\bm{q}, T; t^\prime) \cdot d\bm{S},
\label{eq:flux}
\end{equation}
where $\bm{J}(\bm{q}, T; t^\prime)$ denotes the current Eq.~(\ref{eq:current}). Note that the current $\bm{J}(\bm{q}, T; t^\prime)$ has only components in the $\beta_2$ and
$\beta_3$ directions.
The yield for the fission fragment with mass $A$ is defined by 
\begin{equation}
Y(A) \propto \sum_{\xi \in \mathcal{A}} \lim_{t \rightarrow \infty} F(\xi;t).
\end{equation}
 The set $\mathcal{A}(\xi)$ contains all elements belonging to the scission hyper-surface such that one of 
the fragments has mass number $A$.

\section{\label{sec:results}Illustrative calculation: induced fission dynamics of $^{228}$Th}
As an illustrative example, we have performed a TDGCM calculation of induced fission of $^{228}$Th. 
For this nucleus the charge distribution of fission fragments exhibits a coexistence of symmetric and asymmetric peaks~\cite{Schmidt2000_NPA665-221}.
In the first step, a large scale MDC-RMF calculation is performed to generate the potential energy surface, 
single-nucleon wave functions and occupation factors in the $(\beta_2,\beta_3,T)$ space.  
The intervals for the collective variables are:  $-1 \le \beta_2 \le 7$ with   
a step $\Delta \beta_2 = 0.04$; $0 \le \beta_3 \le 3.5$ with a step $\Delta \beta_3 =0.05$;
and the temperature is varied in the range $0 \le T \le 2.0$ MeV, with a step $\Delta T = 0.1$ MeV.

\begin{figure}[!]
\centering
 \includegraphics[width=0.45\textwidth]{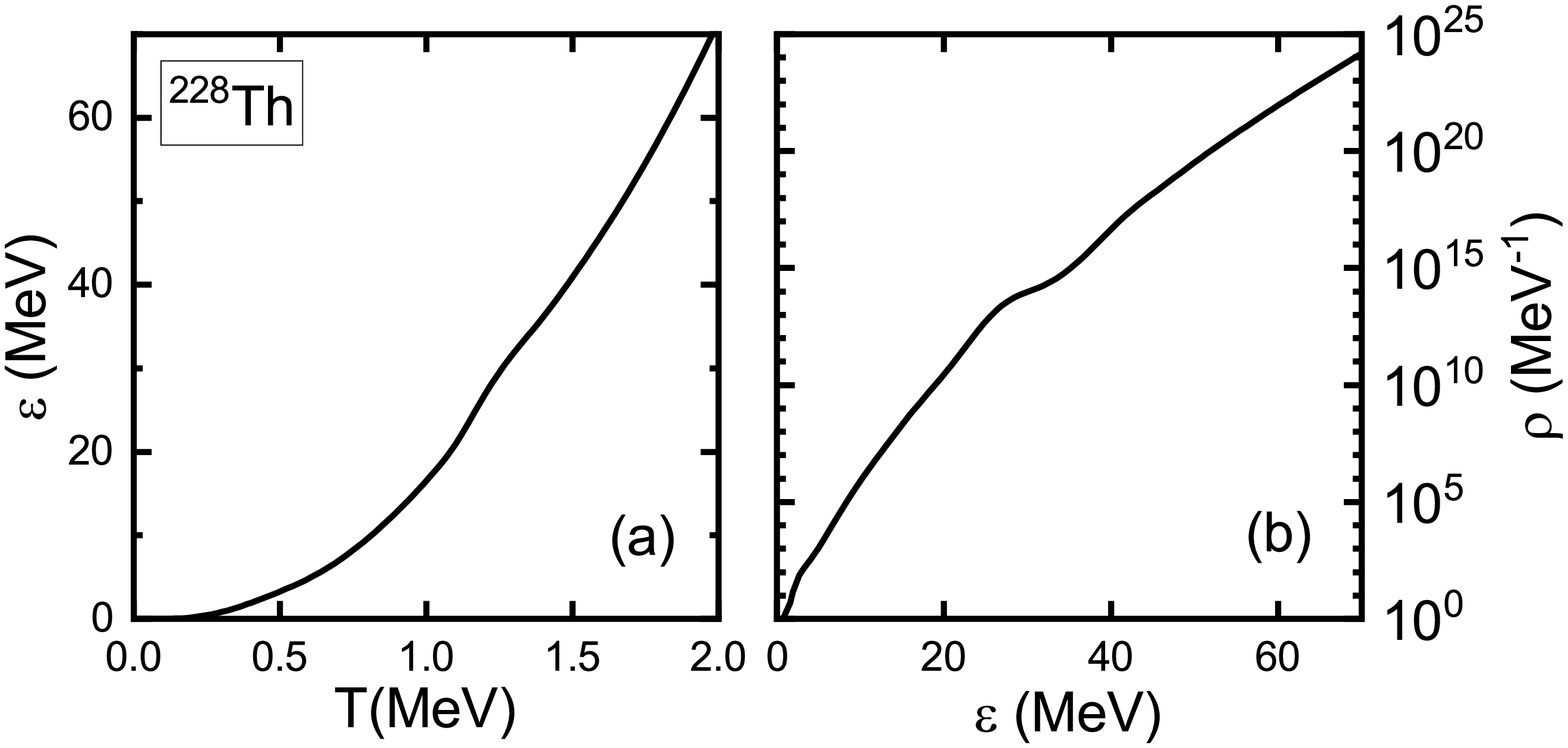}
 \caption{(Color online)~\label{fig:e_ld}%
 Excitation energy of the equilibrium (global) minimum $\epsilon$ as a function of temperature $T$ (a),  
 and the intrinsic level density $\rho$ as a function of excitation energy (b) for $^{228}$Th,
 calculated with the finite-temperature RMF+BCS model based on the DD-PC1 functional and finite-range separable pairing.
}
\end{figure}

Panel (a) in Fig.~\ref{fig:e_ld} displays the calculated excitation energy of the equilibrium minimum $\epsilon$
as a function of temperature $T$. The excitation energy increases quadratically with 
temperature, in accordance with the Bethe formula $\epsilon = aT^{2}$. 
The intrinsic level density, shown in panel (b), increases exponentially with the entropy and, 
therefore, also with excitation energy. 
The change in slope of the excitation energy and the intrinsic level density can be associated with 
the pairing phase transition at the critical temperature $T_{c} \sim 0.7$ MeV, and a shape transition 
at $T \sim 1.2$ MeV.

\begin{figure}[!]
\centering
 \includegraphics[width=0.45\textwidth]{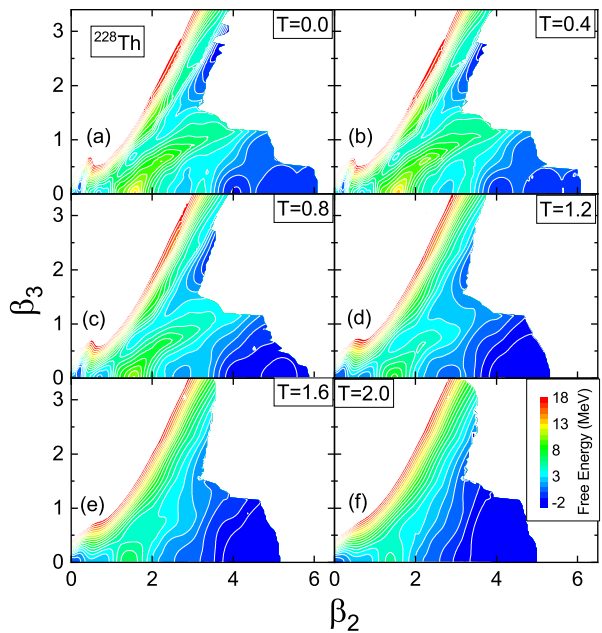}
 \caption{(Color online)~\label{fig:3DPES}%
Free energy manifold of $^{228}$Th, calculated with the RMF+BCS model based on the functional DD-PC1 and finite-range separable pairing, projected 
on the quadrupole-octupole axially symmetric plane, for different values of the temperature $T$. 
In each panel energies are normalized with respect to the corresponding value at the equilibrium minimum.
Contours join points on the surface with the same energy, and the separation between neighboring contours is 1 MeV.
}
\end{figure}
The two-dimensional deformation free energy surfaces $F(T) = E(T) - TS$, in the collective space $(\beta_{2}, \beta_{3})$ 
for selected values of temperature $T=0, 0.4, 0.8, 1.2, 1.6$, and $2$ MeV, are shown in Fig.~\ref{fig:3DPES}.
Only configurations with $\hat{Q}_{N} \geq 3$ are displayed, and the frontier of this domain corresponds to the scission contour. 
The deformation surfaces for $T=0$ and $0.4$ MeV are very similar, with a pronounced ridge separating the asymmetric and symmetric fission
valleys. For temperatures $T \geq 0.8$ MeV this ridge gradually disappears.
 For $T = 0$ MeV, the scission contour starts from an elongated symmetric point at  $\beta_{2} \sim 6$,
and evolves to a minimal elongation $\beta_{2} \sim 3$ as reflection asymmetry increases.
For higher temperatures we notice minor qualitative modifications of the scission contour. 
\begin{figure}[!]
\centering
 \includegraphics[width=0.45\textwidth]{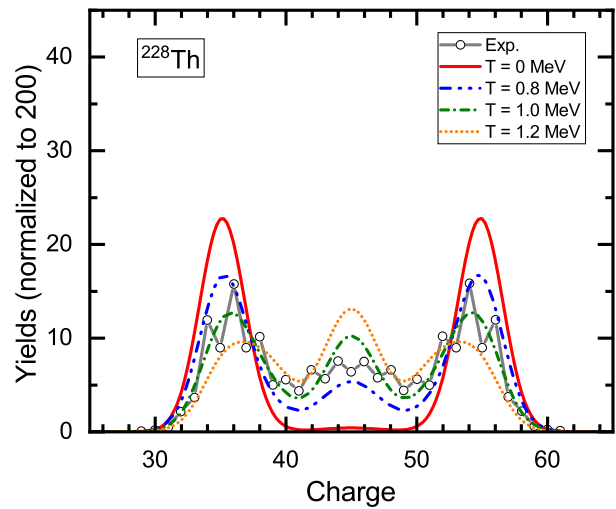}
 \caption{(Color online)~\label{fig:2D_yields}%
Charge yields for induced fission of $^{228}$Th at fixed temperatures: T=0, 0.8, 1.0 and 1.2 MeV.
The average excitation energy of the initial state $E_{\rm coll}^*$ is chosen 1 MeV above the fission barrier. 
 The data for photo-induced fission correspond to photon energies in the interval $8-14$ MeV, 
 and a peak value of $E_{\gamma} = 11$ MeV~\cite{Schmidt2000_NPA665-221}.
}
\end{figure}

 To begin with, we have performed a 2D TDGCM+GOA calculation at fixed temperatures $T=0$, $0.8$, $1.0$
and $1.2$ MeV (for a detailed description of the model, we refer the reader to Ref.~\cite{Zhao2019_PRC99-014618}). This calculation does not include dissipation effects. For each temperature, the average excitation energy of the initial state $E_{\rm coll}^*$ is chosen 1 MeV above the fission barrier. The resulting charge yields, normalized to  $\sum_A{Y(A)}=200$,
are displayed in Fig.~\ref{fig:2D_yields} in comparison to the experimental fragment charge distribution for photo-induced fission of $^{228}$Th with photon energies in the interval $8-14$ MeV \cite{Schmidt2000_NPA665-221}. 
The fission yields are obtained by convoluting the raw 
flux with a Gaussian function of the number of particles, and the width is set to 1.6 units.
For $T=0$ MeV the calculation 
predicts asymmetric peaks located at $Z=35$ and $Z=55$, that is, one charge unit away from the experimental values
$Z=36$ and $Z=54$. Furthermore, the zero temperature calculation fails to describe the empirical yields for symmetric
fission, and overestimates the asymmetric peaks. Increasing the temperature to $T=0.8$ MeV brings the theoretical
charge yields in fair agreement with the experimental values. For even larger values of temperature $T=1.0$ MeV and
$T=1.2$ MeV, the description of charge yields deteriorates. In particular, compared to data the symmetric peak is more pronounced, while
the asymmetric peaks are underestimated and their position is shifted towards the symmetric peak. This is consistent
with the behavior of the free energy surfaces displayed in Fig.~\ref{fig:3DPES}, where the ridge between symmetric and
asymmetric fission valleys decreases with temperature for $T\ge 0.8$ MeV.

\begin{figure}[]
\centering
 \includegraphics[width=0.45\textwidth]{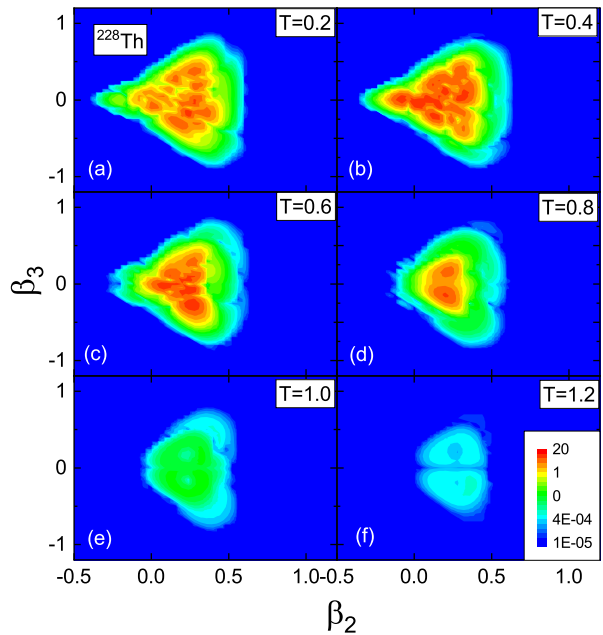}
 \caption{(Color online)~\label{fig:Th228_iniwf}%
 2D projections on the ($\beta_2$, $\beta_3$) plane of the probability distribution of the initial wave packet for 
 induced fission of $^{228}$Th, at different temperatures. The average excitation energy of the initial state is  $E_{\rm coll}^* = 11$ MeV.
}
\end{figure}

\begin{figure}[!]
\centering
 \includegraphics[width=0.45\textwidth]{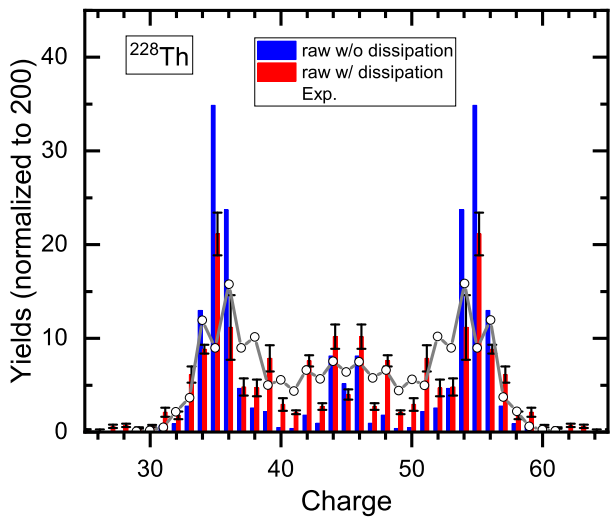}
\caption{(Color online)~\label{fig:yields_raw}%
 Raw charge yields (normalized to 200) for induced fission of $^{228}$Th, 
 calculated in the 3D space of axial deformation parameters $\beta_{2}$, $\beta_{3}$, and temperature $T$. 
 The yields obtained without (blue bars) and with (red bars) dissipation are shown in comparison with available data.
 The results obtained with the inclusion of the dissipation term, correspond to the mean value of five  calculations with different random matrices $\bm{\eta}(T, T^{\prime})$. The resulting 
 standard deviations are shown as error bars.
 The data for photo-induced fission correspond to photon energies in the interval $8-14$ MeV, 
 and a peak value of $E_{\gamma} = 11$ MeV~\cite{Schmidt2000_NPA665-221}.
}
\end{figure}

\begin{figure}[!]
\centering
 \includegraphics[width=0.45\textwidth]{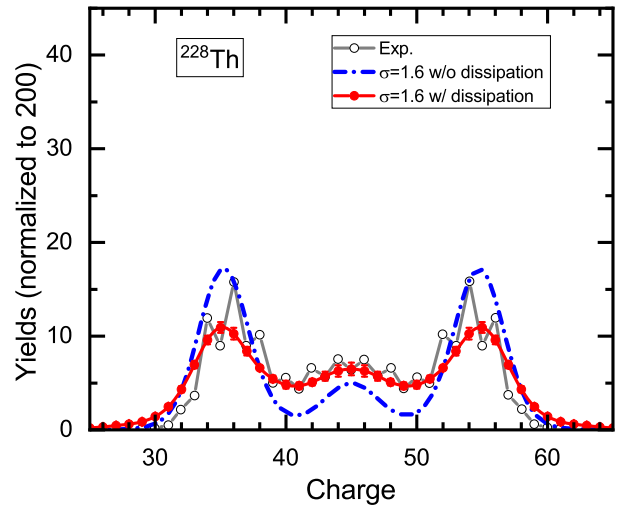}
\caption{(Color online)~\label{fig:yields_sig}%
Same as in the caption to Fig.~\ref{fig:yields_raw} but for the charge yields obtained by convoluting the raw flux with a Gaussian function of width $\sigma = 1.6$ charge units.
}
\end{figure}

In the next step, a full 3D calculation of induced fission dynamics of
$^{228}$Th is carried out in the space of the axial shape variables ($\beta_2$, $\beta_3$), and the temperature T. In Fig.~\ref{fig:Th228_iniwf}, the 2D projections on the $\beta_2$ - $\beta_3$ plane of the probability distribution of the initial wave packet are plotted for different temperatures. The average excitation energy of the initial state is  $E_{\rm coll}^* = 11$ MeV. 
In Figs.~\ref{fig:yields_raw} and~\ref{fig:yields_sig}, we compare the theoretical predictions for the charge yields 
with  the  data for photo-induced fission of $^{228}$Th~\cite{Schmidt2000_NPA665-221}. 
The raw charge yields, normalized to 200, obtained directly from the collective flux through the scission hypersurface are 
displayed in Fig.~\ref{fig:yields_raw}, while in Fig.~\ref{fig:yields_sig} we show the convolution of the raw yields with a Gaussian function of width $\sigma = 1.6$ charge units.
Results obtained without the inclusion of the dissipation term proportional to $\bm{\eta}(\bm{q}; T, T^{\prime})$ in Eq.~(\ref{eq:tdgcmgoa_ft}), are denoted by blue bars in
Fig.~\ref{fig:yields_raw}, and the blue dot-dashed curve in Fig.~\ref{fig:yields_sig}. 

When the dissipation term is included in the Hamiltonian of Eq.~(\ref{eq:tdgcmgoa_ft}) for the statistical wave function, the results denoted by red bars in Fig. ~\ref{fig:yields_raw},  
and the red solid curve in Fig.~\ref{fig:yields_sig}, are obtained. Since the matrix elements of the dissipation function $\bm{\eta}(\bm{q}; T, T^{\prime})$ Eq.~(\ref{eq:eta}) are assumed to be 
Gaussian random variables, the calculation has been carried out with five different random matrices 
$\bm{\eta}(T,T^{\prime})$. The results denoted mean values, and the corresponding  
 standard deviations are shown as error bars. Three values of the strength parameter of the dissipation function have been used in the calculation: $\gamma = 0.001$,  0.01, and 0.05.  The best agreement with data is obtained for $\gamma = 0.01$.

\begin{figure}[!]
\centering
 \includegraphics[width=0.45\textwidth]{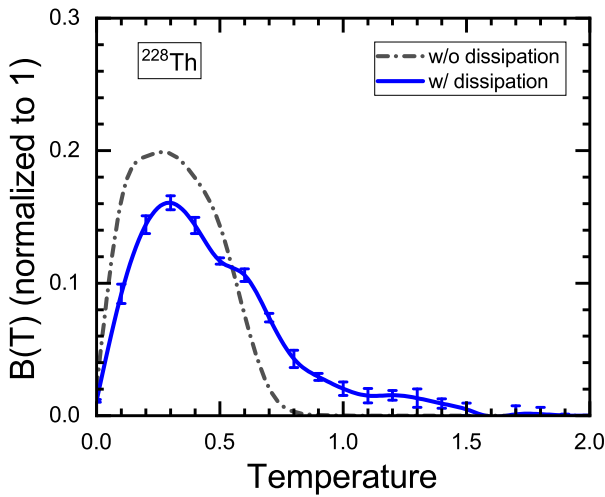}
\caption{(Color online)~\label{fig:yT}%
Time-integrated collective flux $B(T)$ Eq.~(\ref{eq:yT}) through the scission contour, as a function of  temperature. See caption to Fig.~\ref{fig:yields_raw} for the explanation of error bars. 
}
\end{figure}

The results displayed in Figs.~\ref{fig:yields_raw} and \ref{fig:yields_sig} demonstrate that, even though the results obtained without dissipation qualitatively reproduce the trend of the data, it is only when dissipation effects are included that experimental charge yields can be reproduced on a quantitative level. We note that a single parameter has been adjusted to experimental results. To illustrate the effect of dissipation on the flux of the probability current through the scission hyper-surface, in 
Fig.~\ref{fig:yT} we plot the time-integrated flux through the scission contour in the $\beta_{2}$ - $\beta_{3}$ plane, 
for a given value of the temperature $T$  
\begin{equation}
 \label{eq:yT}
B(T) \propto \sum_{\xi \in \mathcal{B}}{\lim_{t\to\infty}F(\xi;t)}.
\end{equation}
The set $\mathcal{B}(\xi \equiv \beta_2, \beta_3)$ contains all elements of the scission contour with a  given value $T$. Without dissipation, a parabolic dependence on temperature is obtained for the time-integrated flux through the scission contour. The parabola, with the maximum at $T \approx 0.3$ MeV, is consistent with the 
the probability distribution of the initial wave packet shown in Fig.~\ref{fig:Th228_iniwf}. When dissipation is included, only the low-T part of the time-integrated flux exhibits a parabolic structure with a maximum at essentially the same temperature. The high-T branch, however, in this case extends to much higher temperatures. While without dissipation no flux through the scission contour is obtained for $T > 0.8$ MeV, 
dissipation broadens the distribution of the flux and the high-T tail reaches $T \approx 1.5$ MeV. The time-integrated flux nicely illustrates how the inclusion of dissipation in the time-evolution of the collective wave function can describe the heating of the fissioning nucleus in the final, saddle to scission phase of fission dynamics. 

%
\section{\label{sec:summary}Summary}
Over the last decade various implementations of the time-dependent generator coordinate method (TDGCM) have successfully been applied in the description of the fission process and, in particular, in the calculation of fission yields \cite{younes19}. However, since TDGCM only includes collective degrees of freedom in the adiabatic approximation, standard formulations of this method cannot take into account the dissipation of the energy of collective motion into intrinsic degrees of freedom. Dissipation heats up the fissioning system, can modify the fission path in the collective space, and affects the time it takes for a nucleus to reach scission \cite{nix83}.

Starting from a quantum theory of dissipation for nuclear collective motion \cite{Kerman1974_PS10-118}, in this work we have extended the temperature-dependent TDGCM for induced fission dynamics, to allow for  dissipation effects. The extension is based on a generalization of the GCM generating functions that includes excited intrinsic states. Because the level density is high even at relatively low excitation energies, the discrete label of excited intrinsic states is replaced by a continuous energy variable. By introducing a statistical wave function defined by the average value of the collective wave function and the level density, an equation of motion is derived in the collective coordinates and excitation energy. With the usual assumption that the Hamiltonian overlap kernel decreases rapidly with distance between the corresponding collective coordinates, an expansion in a power series in collective momenta leads to a Schr\"odinger-like equation that explicitly includes a dissipation term, proportional to the momentum of the statistical wave function. By expressing the excitation energy in terms of nuclear temperature, the new model can be formulated in the framework of temperature-dependent TDGCM, in which the Helmholtz free energy plays the role of the collective potential, and the collective inertia is calculated in the finite-temperature perturbative cranking approximation. 

An illustrative calculation has been performed for induced fission of $^{228}$Th, for which  
the charge distribution of fission fragments exhibits a coexistence of symmetric and asymmetric peaks. The three-dimensional model space includes the axially-symmetric quadrupole and octupole shape variables, and the nuclear temperature. The corresponding dissipation function, even though in principle it could be determined microscopically, is approximated by a set of Gaussian random variables, effective in the region of the collective space between the second fission barrier and the scission hyper-surface. When compared to data for photo-induced fission of $^{228}$Th \cite{Schmidt2000_NPA665-221}, the calculated fission yields clearly demonstrate the important role of the dissipation term in the hamiltonian for the statistical wave function. Even though already the finite-temperature TDGCM without dissipation reproduces the empirical trend of the data, it is only with the explicit inclusion of dissipation that theoretical results are obtained in quantitative agreement with the experimental charge yields. As a result of dissipation, a high-temperature tail broadens the distribution of the flux through the scission hyper-surface as function of temperature. 

The TDGCM extended with the explicit inclusion of dissipation, and the results of the pilot study of induced fission dynamics reported in this work, point to a new microscopic approach that can be employed to quantitatively describe dissipation of the energy of collective motion into intrinsic excitations. Obviously, the next step is to go beyond the statistical ansatz for the dissipation function, and derive it microscopically from the underlying hamiltonian. An important problem is also the definition of the scission hyper-surface in a multi-dimensional space that includes temperature or intrinsic excitation energy. The interplay between dissipation and pairing degrees of freedom should be explored, as well as the effect of dissipation on the total kinetic energy distribution.   
\bigskip
\acknowledgements
This work has been supported in part by the QuantiXLie Centre of Excellence, a project co-financed by the Croatian Government and European Union through the European Regional Development Fund - the Competitiveness and Cohesion Operational Programme (KK.01.1.1.01.0004) and the Croatian Science Foundation under the project Uncertainty quantification
within the nuclear energy density framework (IP-2018-01-5987).
It has also been supported by the National Natural Science Foundation of China under Grant No. 12005107 and No. 11790325.
The authors acknowledge the Beijing Super Cloud Computing Center (BSCC) for providing HPC resources 
that have contributed to the research results reported within this paper \url{URL: http://www.blsc.cn/}.


\end{document}